\RequirePackage{fix-cm}
\documentclass[smallextended]{svjour3}       
\smartqed  
\usepackage{graphicx}
\begin{document}

\title{Power Grid with 100\% Renewable Energy for Small Island Developing States
\thanks{The present study was supported by the Ministry of Education, Science, Sports, and Culture, Grants-in-Aid for Scientific Research (B), Grant No. 17KT0034
(2017-2019) and Exploratory Challenges on Post-K computer (Studies of Multi-level Spatiotemporal Simulation of Socioeconomic Phenomena).}
}
\subtitle{Nexus of Energy, Environment, and Economic Growth}


\author{Yuichi Ikeda         
}


\institute{Yuichi. Ikeda \at
              Graduate School of Advanced Integrated Studies in Human Survivability,
Kyoto University \\ 
              Tel.: +81-75-762-2102\\
              \email{ikeda.yuichi.2w@kyoto-u.ac.jp}   \\
             \emph{1 Yoshida-Nakaadachi-cho, Sakyo-ku, Kyoto-shi, Kyoto 606-8306, JAPAN}
}

\date{Received: date / Accepted: date}

\maketitle

\begin{abstract}
We estimated system-wise levelized cost of electricity (LCOE) for a power grid with a high level of  renewable energy using our grid optimization model.
The estimation results of the system-wise LCOE are discussed in terms of the nexus of energy, environment, and economic growth for Small Island Developing States (SIDS) economies.
While 100\% renewable energy is technologically possible with the usage of electricity storage, the estimated LCOE is as high as 397 \$/MWh which is substantially higher than electricity prices for residential consumers in the US and Japan.
The susceptibility analyses imply that the estimated LCOE increase of 223\% with a 100\% renewable power grid corresponds to an as high as 11\% decrease in economic growth. 
This decrease in economic growth would have a significant negative impact on SIDS economies.
However, hydrogen production via the electrolysis of water using the excess energy supply from solar photovoltaics would reduce the LCOE, therefore higher economic growth would be attained with less $\rm{CO}_2$ emission.
\keywords{Energy \and Power grid \and Renewable energy \and Levelized cost of electricity \and Economic growth}
\end{abstract}

\section{Introduction}
\label{intro}

The population size of Small Island Developing States (SIDS) is rather limited. SIDS have a combined population of approximately 65 million \cite{Ohrlls2015}, which is approximately 1 \% of the world's population. In SIDS, nearly 30 \% of the population lives at elevations of less than 5 m above sea level. Therefore, SIDS are often said to be the most vulnerable areas to climate change due to the currently increasing $\rm{CO}_2$ emissions. 

The Intergovernmental Panel on Climate Change (IPCC) has reported that emissions resulting from human activities are substantially increasing atmospheric concentrations of greenhouse gases, resulting on average in additional warming to the Earth's surface \cite{IPCC1990}. Based on IPCC reports, policy makers in various countries, including advanced countries, emerging economies, and SIDS, have proposed energy policies to introduce as much renewable energy as possible to curtailing $\rm{CO}_2$ emissions. Therefore renewable energy is considered as a key to curtail $\rm{CO}_2$ emission. However, skepticism persists due to the high cost of investment in renewable energy and its integration into power grids. 

Here, we briefly review the sustainable development goals (SDGs) \cite{Sachs2015}  and discuss the synergy and trade-off between these goals. 
The SDGs consists of 17 goals and 169 targets. Some targets are common to different goals.
Le~Blanc expressed the SDGs as a bipartite network between the 17 goals and 169 targets \cite{Blanc2015}.
Contracting the bipartite network we obtain a network that only consists of targets.
The links of the obtained network are weighted, even though the links of the original bipartite network are  binary.
Many high degree nodes are located at the center of the network, e.g., 
SDG 1 ``poverty'', SDG 10 ``inequality'', SDG 12 ``sustainable consumption and production'', and 
SDG 8 ``growth and employment''.
The synergy between SDGs is identified as pairs of nodes connected by large weight links, e.g.,
SDG 1 ``poverty'' and SDG 10 ``inequality'', SDG 5 ``gender'' and SDG 4 ``education'', and
SDG 16 ``peaceful and inclusive'' and SDG 10 ``inequality''.
On conversely, trade-off between SDGs is identified as pairs of nodes without link, e.g., 
SDG 7 ``energy'' and SDG 13 ``climate change''.

Here, we point out the need for a concrete example studying the trade-off between energy goals and climate change. 
In this paper, we consider a power grid with high renewable energy in SIDS economy, and discuss nexus of energy, environment and economic growth.
The goal of this paper is to estimate the electricity price for a power grid with high renewable energy for a SIDS economy and to discuss how the estimated electricity price affects the economic growth.
This paper is organized as follows. 
In Section \ref{sec:2}, a model of grid integration of renewable energy is described.
In Section \ref{sec:3}, the detail of our analysis is explained.
In Section \ref{sec:4}, various results are shown and discussion on nexus of energy, environment and economic growth for a SIDS economy is given.
Section \ref{sec:5} summarizes the paper.

\section{Grid Integration of Renewable Energy}
\label{sec:2}

Our model of grid integration for renewable energy is described in this section \cite{Ikeda2014} \cite{Ikeda2015} \cite{Ikeda2018}.
We assume here that the thermal power plants in a SIDS economy consists primarily of diesel power, i.e., less coal and natural gas thermal power. 
We also assume that the renewable energy in SIDS economy is mainly solar photovoltaic (PV), i.e., less wind power, because the low altitude area where most SIDS are located are rich in solar potential and poor in wind potential \cite{Chatzivasileiadis2013}.

\subsection{Power Grid}
\label{subsec:GridIntegration}

A typical conventional power grid today includes thermal power plants on the supply side and also factories, office buildings, and individual residences on the demand side. The demand varies and fluctuates with time. This fluctuation needs to be balanced by changing the output of the thermal power plants.

The evolution of power grids is currently underway and moving toward the 1st phase of smart grid. 
Solar PV panels are starting to be installed in various locations on power grids on both the supply and demand sides. The installation of renewable energy introduces additional fluctuation into a power grid. This requires additional balancing power on the supply side, even though the demand for electricity on the supply side decreases due to the additional capacity of the renewable energy. Therefore, we need a certain capacity of electricity storage to obtained sufficient balancing power for the stable operation of the power grid. Currently, the cost of electricity storage is high: therefore, the price of electricity must increase. 

In the near future, demand side management systems which consist of a smart meter, a home energy management server, and appliances applicable to the server, will become popular and we will obtain a new source of balancing power on the demand side. This new balancing power will reduce the capacity of the electricity storage required to balance fluctuations in the renewable energy.
We call this the 2nd phase of the smart grid. However it might be delayed to diffuse the technology of the 2nd phase of smart grid, due to the complicated relationships between  power grid stakeholders. 

Our grid model aims to describe the 1st phase of a smart grid in order to estimate the electricity price for a power system with a high level of renewable energy integration in a SIDS economy.

\subsection{Grid Model}
\label{subsec:Optimization}

\begin{figure*}
  \includegraphics[width=0.80\textwidth]{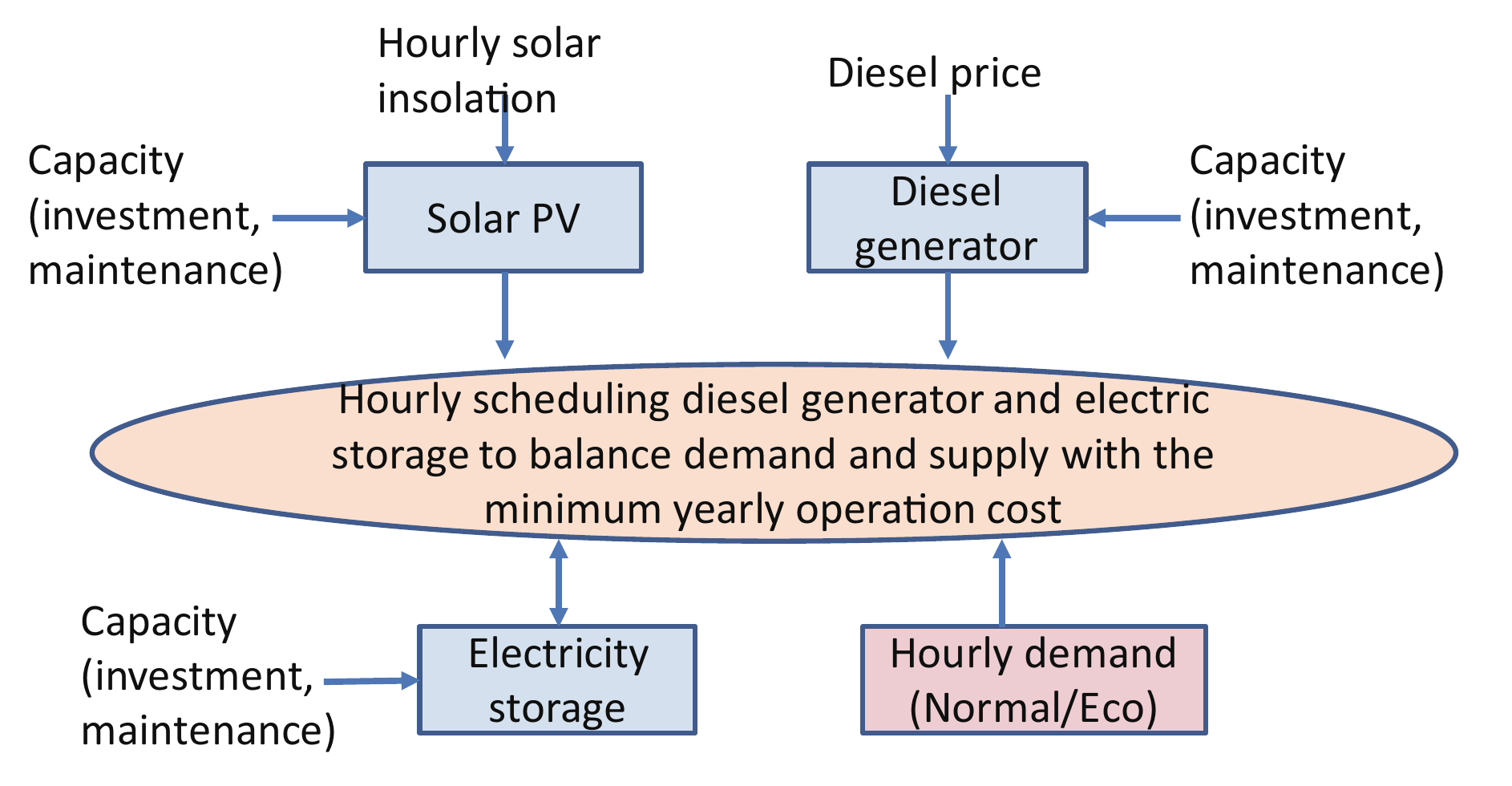}
\caption{Concept of our grid model. The model is formulated using Eqs. (\ref{Eq:Objective}) - (\ref{Eq:LCOE}).}
\label{fig:1} 
\end{figure*}

We formulate our grid model as an optimization model.
The concept of our grid model is shown in Fig. \ref{fig:1}.
First, we define the objective function which is equal to the fuel expenditure required to operate the thermal power plants:
\begin{equation}
 \sum_{t=1}^{T} \sum_{i=1}^{N} b_i p_t^i,
\label{Eq:Objective}
\end{equation}
where $b_i$ is the fuel cost used to obtain a unit of electricity from a thermal power plant $i$ and $p_t^i$ is the output electricity at a time $t$ from the plant $i$. $T$ and $N$ are the time period of the calculation and the number of thermal power plants, respectively. In this paper, the time period is one year and the time step is one hour, therefore $T=24 \times 365 = 8760$.

The objective function in Eq. (\ref{Eq:Objective}) is minimized under the global constraints of the demand-supply balance and some local constraints on the electricity storage. The global constraints of the demand-supply balance are given as
\begin{equation}
 \frac{\sum_{i=1}^N p_t^i + pv_t^{(f)} + g_t - h_t - d_t^{(f)}}{\sqrt{\sigma_d^2 + \sigma_p^2}} \ge \phi^{-1} (\alpha),
\label{Eq:Constraint}
\end{equation}
where $pv_t^{(f)}$, $g_t$, $h_t$, $d_t^{(f)}$, $\sigma_d^2$, and $\sigma_p^2$ are the solr PV output forecast at $t$,  the discharged electricity from the electricity storage at $t$, the electricity charged to storage at $t$, the demand forecast at $t$, the variance of the demand fluctuation, and the variance of the fluctuation in the solar PV output, respectively.
Note that the fluctuations of the demand and renewable energy are stochastically taken into account.
$\alpha$ and $\phi(\alpha)$ are the probability of ensuring the supply-demand balance and the cumulative distribution function, respectively.
In this paper we use $\alpha=1.28$ which means that the demand-supply balance is satisfied with a probability of $90\%$.
The cumulative distribution function is written using the error function $erf[\cdot]$ as
\begin{equation}
 \phi (x) = \frac{1}{2} \left( 1 + erf \left[ \frac{x-\mu}{\sqrt{2 \sigma^2}} \right] \right).
\label{Eq:Cumulative}
\end{equation}
Here $\mu=0$ and $\sigma=1$, because the left hand side of Eq. (\ref{Eq:Constraint}) is normalized.
The local constraints on the electricity storage are given as
\begin{equation}
 v_t c_{min} \le g_t \le v_t c_{max},
\label{Eq:Storage1}
\end{equation}
\begin{equation}
 (1-v_t) c_{min} \le h_t \le (1-v_t) c_{max},
\label{Eq:Storage2}
\end{equation}
\begin{equation}
 R_{min} \le \sum_{s=1}^t (h_s \eta - g_s) \Delta_t \le R_{max},
\label{Eq:Storage3}
\end{equation}
where $v_t$, $c_{min}$, $c_{max}$, $R_{min}$, $R_{max}$, $\eta$, and $\Delta_t$ are the state variable of the electricity storage ($v_t=1$: discharge, $v_t=0$: charge), minimum discharge power, maximum discharge power, minimum stored energy, maximum stored energy, efficiency, and time
step, respectively.

Next, we describe the system-wise levelized cost of electricity (LCOE).
The system-wise LCOE is estimated by including the thermal power plants, solar PV, and electricity storage.
Therefore the system-wise LCOE is interpreted as a measure of the electricity price, even though the cost of transmission and distribution and the profit of the utility company are not included.
The system-wise LCOE is defined by the aggregated expenditure (\$) to supply a unit amount of electricity (MWh) to the consumer: 
\begin{equation}
 {\rm LCOE} = \frac{\sum_{y=1}^{Y} \frac{I_y + M_y + F_y}{(1+r)^y}}{\sum_{y=1}^{Y} \frac{E_y}{(1+r)^y}},
\label{Eq:LCOE}
\end{equation}
\begin{equation}
 E_y = \sum_{i=1}^N P_y^i + PV_y^{(f)} + G_y - H_y
\label{Eq:LCOE2}
\end{equation}
\begin{equation}
 P_y^i =  \sum_{t=1}^T p_t^i,
\label{Eq:LCOEsub1}
\end{equation}
\begin{equation}
 PV_y^{(f)} = \sum_{t=1}^T pv_t^{(f)},
\label{Eq:LCOEsub2}
\end{equation}
\begin{equation}
 G_y = \sum_{t=1}^T g_t, 
\label{Eq:LCOEsub3}
\end{equation}
\begin{equation}
 H_y = \sum_{t=1}^T h_t, 
\label{Eq:LCOEsub4}
\end{equation}
where $I_y$, $M_y$, $F_y$, $r$, and $T$ are the investment expenditures  (including finance) in year $y$, the operation and maintenance expenditures in year $y$, the fuel expenditures in year $y$, the discount rate, and the life of the system, respectively.
Note that $I_y$ and $M_y$ are aggregated for all of the thermal power plants, renewable energy, and electricity storage on the supply side.
Conversely, $F_y$ is only aggregated for all of the thermal power plants.
We assumes that $P_y^i$, $PV_y^{(f)}$, $G_y$, and $H_y$ are constant during the life of system $Y$.

\section{Analysis Conditions}
\label{sec:3}

We assume a representative case to analyze  a power grid with a high level of renewable energy in a SIDS economy.
The details of the analysis conditions for a representative case are  explained in this section.

\subsection{Representative Case}
\label{subsec:Representative}

We quantitatively estimated electricity prices for power grids in the Temburong area of Brunei with different share of renewable energy using the grid model described in Section \ref{subsec:Optimization}. Even though Brunei is not categorized as a SIDS area in a precise sense, the Temburong area is an isolated enclave of Brunei and its geographical and demographical conditions are similar to many SIDS area. For this reason, we assume that the Temburong area is equivalent to SIDS in terms of power grid analysis.

\begin{figure*}
  \includegraphics[width=0.65\textwidth]{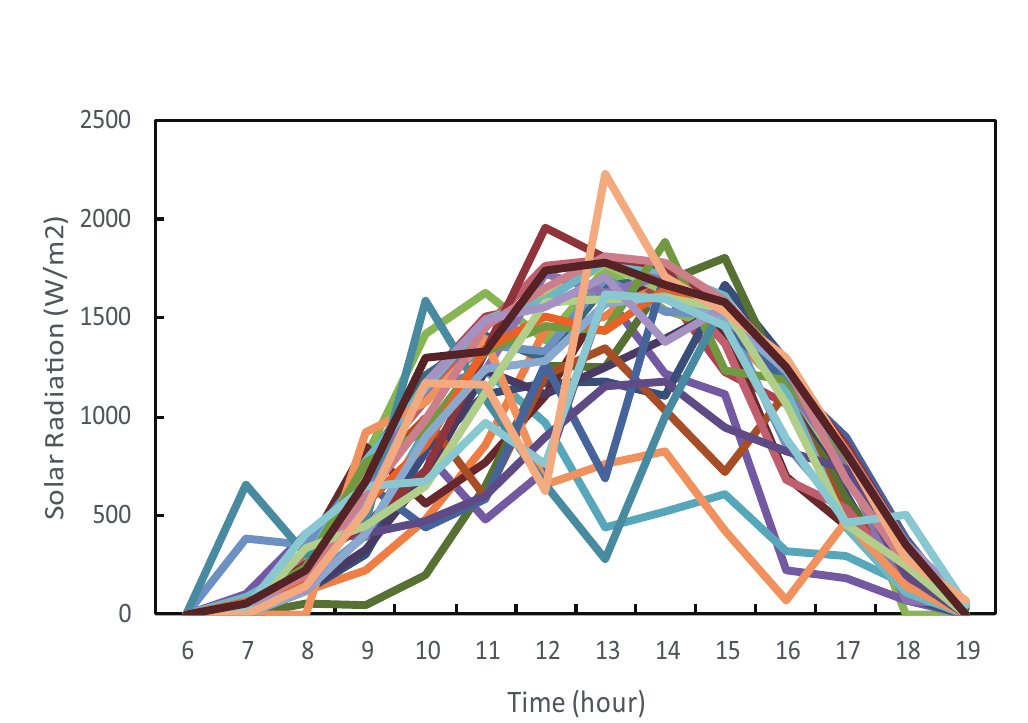}
\caption{Time series of the solar radiation in Brunei. Each line corresponds to the temporal change in the solar radiation from 6:00 am to 20:00 pm in a single day.}
\label{fig:2} 
\end{figure*}

Time series of the solar radiation in Brunei are shown in Fig. \ref{fig:2} for January, 2016.
Each line corresponds to the temporal change in the solar radiation from 6:00 am to 20:00 pm on a single day.
Figure \ref{fig:2} shows that the fluctuation can be considerable depending on the weather conditions, e.g., the coefficient of variation is $28\%$ at noon. 

\begin{figure*}
  \includegraphics[width=0.65\textwidth]{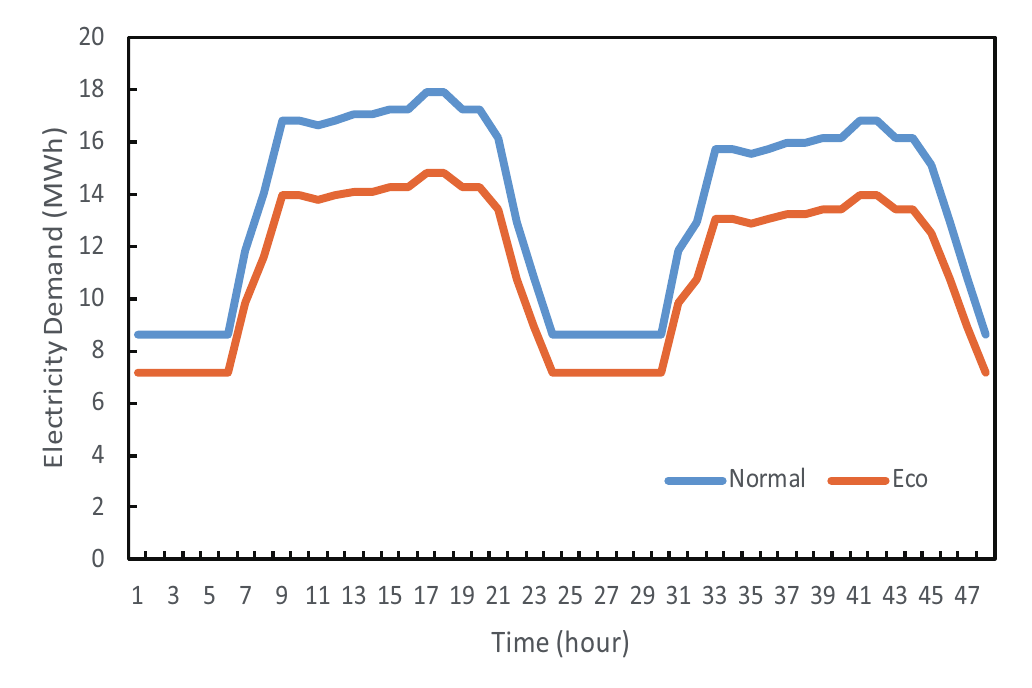}
\caption{Estimated electricity demand for a weekday and a weekend day. The eco-demand is estimated for energy saving buildings. A detailed description of the demand estimation is given in Ref. \cite{Ikeda2018}}
\label{fig:3} 
\end{figure*}

The estimated electricity demand is shown in Fig. \ref {fig:3} for a weekday and a weekend day. The first 24 hours shows the demand for a weekday, and the last 24 hours shows that for a weekend day.
The shape of the estimated electricity demand is fairly flat during the daytime and shows peaks in the early evening.
The demand on a weekend day is considerably lower than the demand on a weekday.   
The blue curve and red curve are the normal demand and the eco-demand, respectively.
Here the eco-demand is estimated for energy-saving buildings. 
The peak of the eco-demand is approximately $78 \%$ that of the normal demand.
A detailed description of this demand estimation is given in Ref. \cite{Ikeda2018}.


\subsection{Parameters on the Supply Side}
\label{subsec:SupplySide}

The parameters for the diesel generators are shown in Table \ref{tab:1}.
The number of diesel generators $N$ changes over the range of $0$ - $7$.
We assume that the diesel fuel price in Brunei is $1/3$ the price in the US.
Note, however, that this assumption might not be valid for all SIDS economies. We simply use this assumption to provide a generic value in this representative case.  
In addition, when estimating the road factor of the diesel generators, we assume that the three-month maintenance is scheduled after one year of generator operation.

The solar PV parameters are shown in Table \ref{tab:2}.
The installed  solar PV capacity varies in the range of $0.0$ - $192.0$ (MW).
Note that the initial investment cost $1500$ \$/kW corresponds to the generation cost $5$ c/kWh. This means that the generation cost of solar PV is not excessive today.
The load factor $0.18$ is relatively high for the solar PV generation. We assume that this value is plausible for most SIDS economies because most SIDS are located at low latitudes.  

The electricity storage parameters are shown in Table \ref{tab:3}.
Here, we consider battery-type electricity storage, such as Li-ion batteries, NaS batteries, lead acid batteries, and vanadium redox flow batteries.
We assume a proportionality between the maximum discharge power $x$ (MW) and the maximum stored electricity $6 \cdot x$ (MWh).
Even though the efficiency $0.7$ varies depending on the technology used for the storage, we use this as a generic value.

For the LCOE estimation in Eq. (\ref{Eq:LCOE}), we assume a life of system of $Y=20$ (years) and a discount rate of $r=0.05$.
Note that the fuel expenditures $F_y$ and the electricity generation $E_t$ calculated by aggregating the results of the optimization of our grid model described in Section \ref{subsec:Optimization} at an hourly time step over one year are used throughout the $20$ years of the life of the system.

\begin{table}
\caption{Parameters of diesel generator}
\label{tab:1} 
\begin{tabular}{lll}
\hline\noalign{\smallskip}
Parameter & Value \\
\noalign{\smallskip}\hline\noalign{\smallskip}
Fuel price (\$/litter) (Diesel price in Brunei, 1/3 of US) & $0.32$ \\
Rated power (MW) & $3.0$ \\
Road factor (Three months maintenance after one year operation) & $0.8$ \\
Initial investment cost (\$/kW) & $650$ \\
Operation and maintenance cost (\$/MW/year)	 & $15 \times 1000$ \\
\noalign{\smallskip}\hline
\end{tabular}
\end{table}
\begin{table}
\caption{Parameters of solar PV}
\label{tab:2} 
\begin{tabular}{lll}
\hline\noalign{\smallskip}
Parameter & Value \\
\noalign{\smallskip}\hline\noalign{\smallskip}
Installed capacity (MW) & $0.0$ to $192.0$ \\
Rated power of unit module (MW) & $6.0$ \\
Road factor & $0.18$ \\
Initial investment cost (\$/kW) & $1500$ (Generation cost 5c/kWh) \\
Operation and maintenance cost (\$/MW/year) & $15 \times 1000$ \\
\noalign{\smallskip}\hline
\end{tabular}
\end{table}
\begin{table}
\caption{Parameters of electricity storage}
\label{tab:3} 
\begin{tabular}{lll}
\hline\noalign{\smallskip}
Parameter & Value \\
\noalign{\smallskip}\hline\noalign{\smallskip}
Minimum discharge power (MW) & $0.0$ \\
Maximum discharge power (MW) & $0.0$ to $95.0$ \\
Minimum electricity storage capacity (MWh) & $0.0$ \\
Maximum electricity storage capacity (MWh) & $0.0$ to $570.0$ \\
Efficiency & $0.7$ \\
Initial investment cost (\$/kWh) & $200$ \\
\noalign{\smallskip}\hline
\end{tabular}
\end{table}

\section{Results and Discussion}
\label{sec:4}

The estimation results for the grid operation and system-wise LCOE using the grid model described in Section \ref{subsec:Optimization} are shown and discussed in terms of the nexus of energy, environment and economic growth for a SIDS economy in this section.

\subsection{Results}
\label{subsec:Results}

\begin{figure*}
  \includegraphics[width=0.80\textwidth]{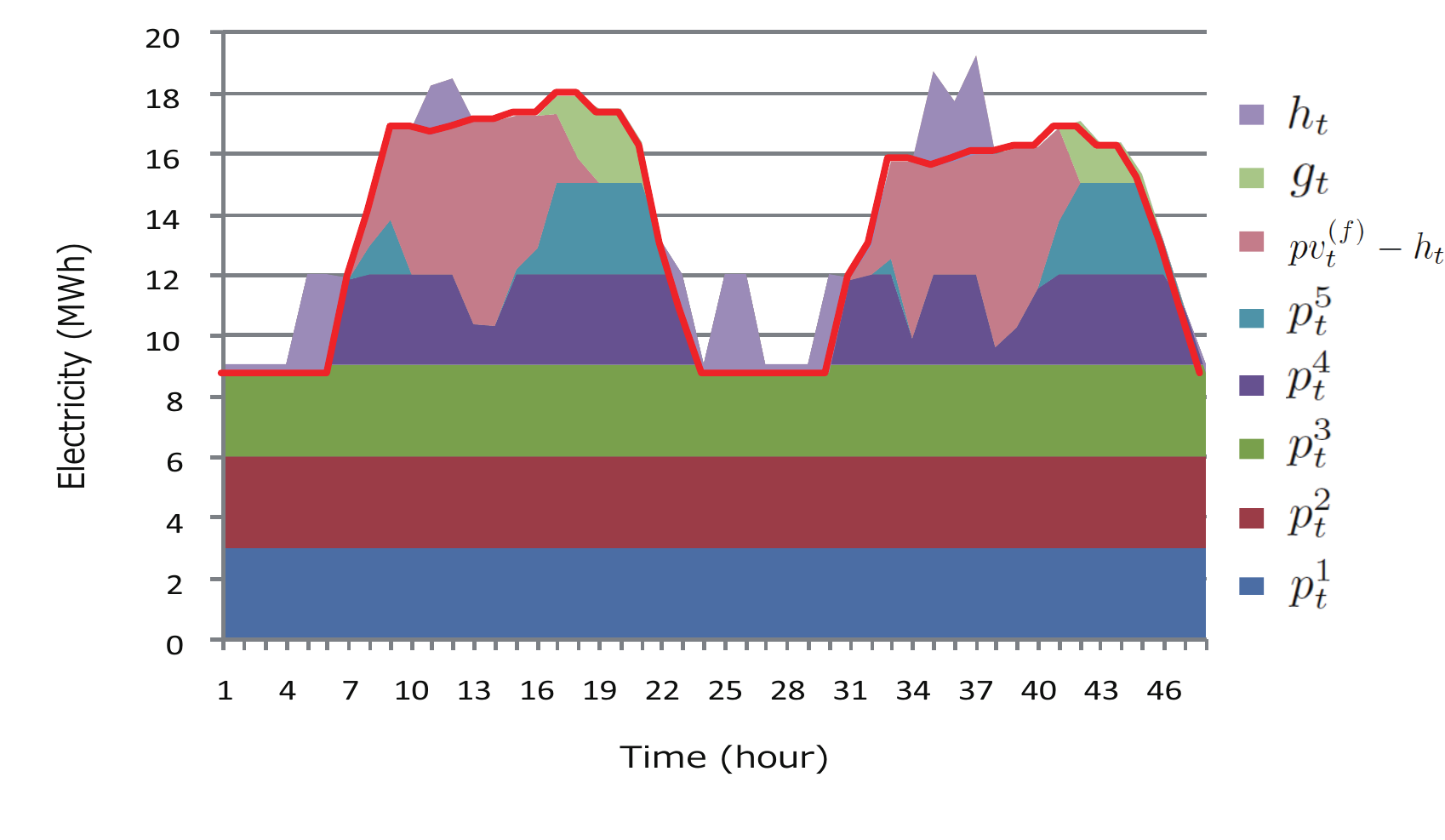}
\caption{Electricity supply on a weekday and a weekend day in January. The electricity demand is indicated by a red curve for a weekday and a weekend day. $p_t^1$ - $p_t^5$ show the outputs of the five diesel generators. $h_t$ shows the excess output from solar PV charged into the electricity storage, and $g_t$ shows the discharged electricity.}
\label{fig:4} 
\end{figure*}

The electricity supply required to satisfy a given demand was estimated throughout a year with a given output from the solar PV of $12$ MW.
Five diesel generators and a maximum electricity storage capacity of $30$ MWh are required on the supply side to satisfy the normal demand throughout a year.
The estimation results for the electricity supply on a weekday and a weekend day in January are shown in Fig. \ref{fig:4}.
Here, the given electricity demand is indicated by a red curve for a weekday and a weekend day.
The number of diesel generators is reduced from seven to five using a maximum electricity storage capacity of $30$ MWh.
In Fig. \ref{fig:4}, $p_t^1$ - $p_t^5$ show the outputs of the five diesel generators, 
$h_t$ shows the excess output from solar PV charged into the electricity storage, and  
$g_t$ shows the discharged electricity.

\begin{figure*}
  \includegraphics[width=0.55\textwidth]{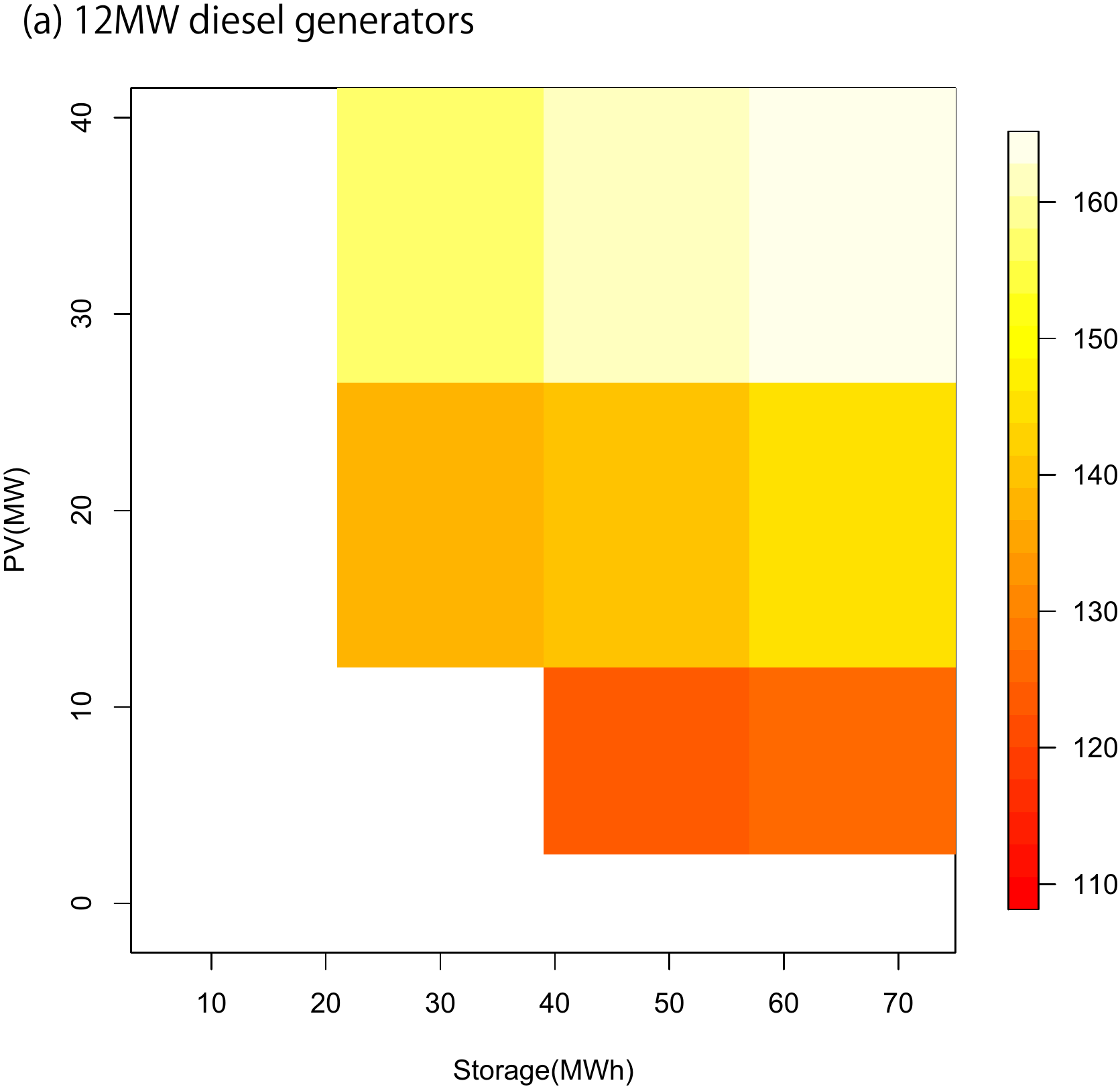}
  \includegraphics[width=0.55\textwidth]{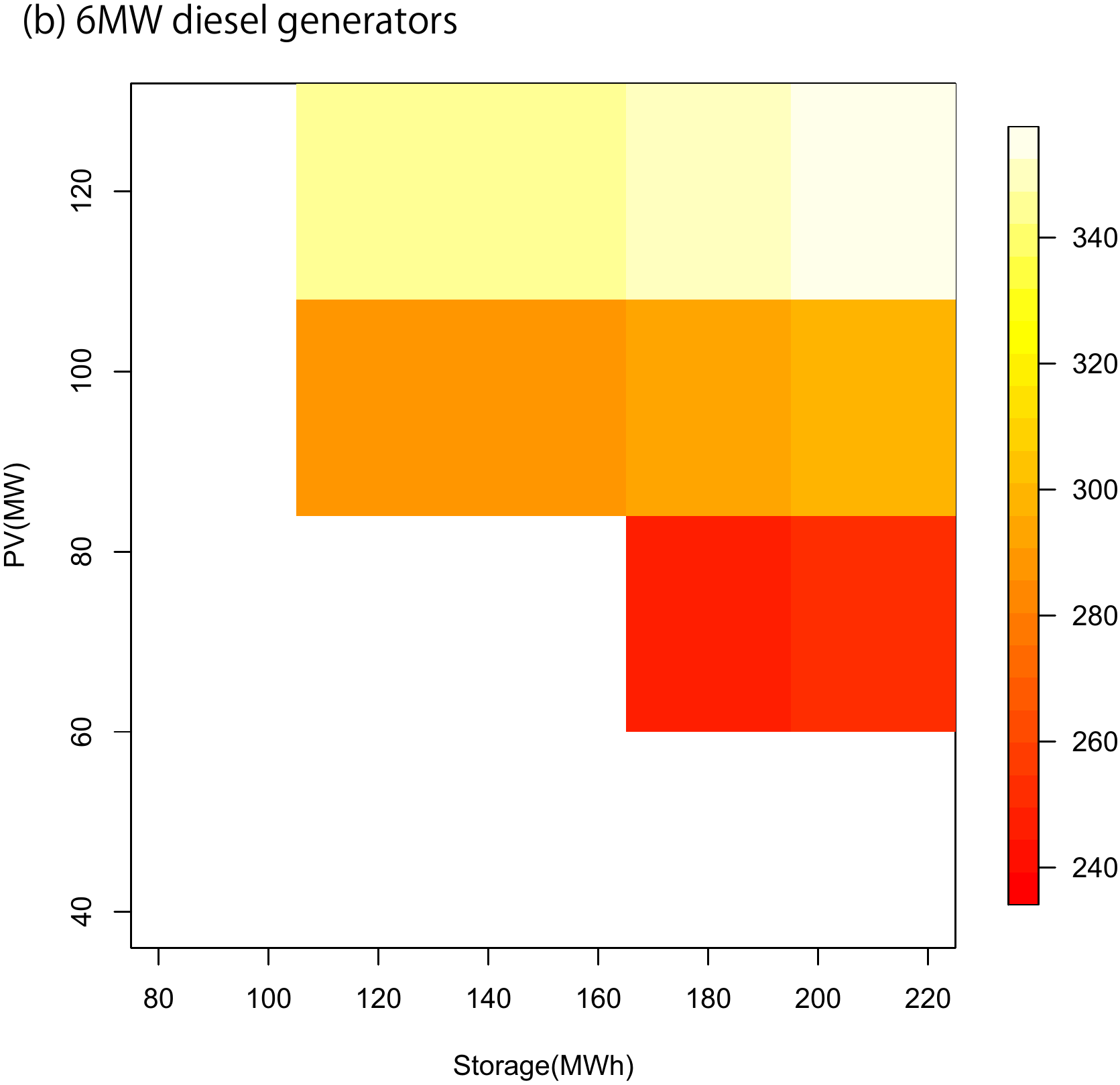}
  \includegraphics[width=0.55\textwidth]{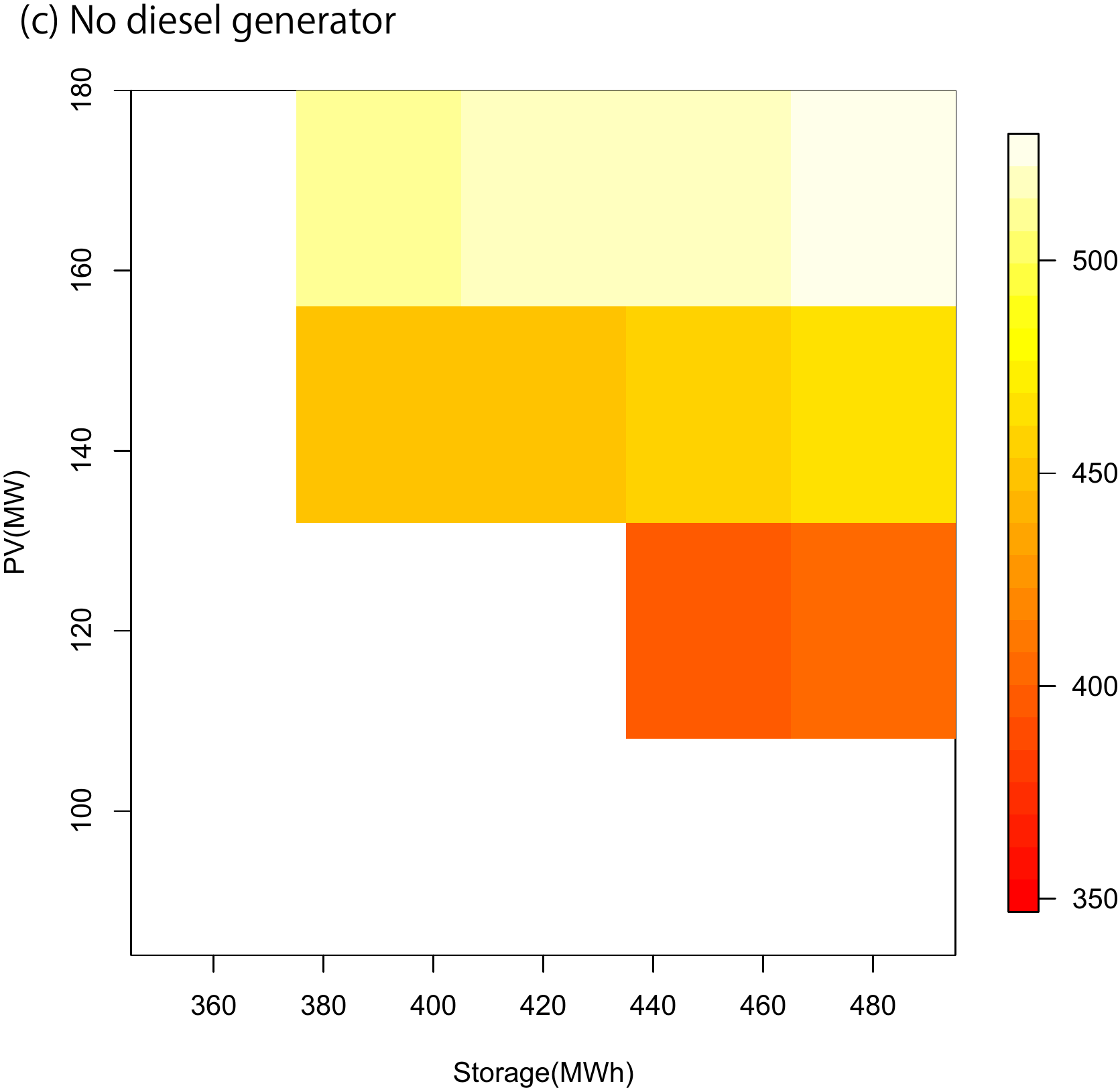}
\caption{System-wise LCOE as a function of the solar PV and electricity storage capacities for the eco-demand case. The white area indicates that the optimization is not feasible for the specified solar PV and the electricity storage capacities. The lowest system-wise LCOEs are 123 \$/MWh, 245 \$/MWh, and 397 \$/MWh in panel (a), (b), and (c), respectively.}
\label{fig:5} 
\end{figure*}

The estimated system-wise LCOEs are shown in Fig. \ref{fig:5} as a function of the solar PV and electricity storage capacities for the eco-demand case. 
The white area in Fig. \ref{fig:5} indicates that the optimization described in Section \ref{subsec:Optimization} is not feasible for the specified solar PV and electricity storage capacities. 
For the systems in panel (a), the installed diesel generator and solar PV capacities are 12 MW and 6 MW, respectively.
The maximum stored electricity is 48 MWh, and the lowest system-wise LCOE is estimated to be 123 \$/MWh.
For the systems in panel (b), the installed diesel generator and solar PV capacities are 6 MW and 72 MW, respectively, the maximum stored electricity is 180 MWh, and the lowest system-wise LCOE is estimated to be 245 \$/MWh.
For the systems in panel (c), the install diesel generator and solar PV capacities are 0 MW and 120 MW, respectively, 
the maximum stored electricity is 450 MWh, 
and the lowest system-wise LCOE is estimated to be 397 \$/MWh.
The 12-MW and 0-MW diesel generator cases correspond to the business-as-usual scenario and the 100\% renewable energy scenario, respectively. 
While 100\% renewable energy, i.e., no diesel generator, is technologically possible with the usage of electricity storage, the estimated LCOE is 397 \$/MWh.
Note that this estimated LCOE is substantially higher than electricity prices for residential consumers in the US (125 \$/MWh) and Japan (253 \$/MWh) \cite{IEA2015}.

\subsection{Discussion}
\label{subsec:Discussion}

The nexus of energy, environment, and economic growth are discussed here to provide invaluable insights toward the realization of SDGs.
The susceptibility analyses of the electricity price on the economic growth can be broken down into two stages. The first stage is an analysis of the relationship between the electricity demand and the economic growth.
The second stage is an analysis of the price-elasticity of the demand for electricity.

Studies in the first stage can be summarized as follows.
\begin{itemize}
  \item In 17 Taiwanese industries, a study of electricity consumption and economic growth indicated that a $1$\% increase in the electricity consumption boosts the real GDP by $1.72$\% \cite{Lu}.
  \item In China, from 1950 to 1980, the elasticity of electricity consumption on economic growth was increased from $2.33$ to $2.72$ which is significantly greater than $1$.
From 1990 to 2013, the growth rate of the electricity consumption was highly correlated with change in the GDP growth \cite{Zhanga}.
  \item In the US, the electricity consumption growth and GDP growth occurred at a similar rate from 1949 to 1996: however, from after 1996 to today, this correlation gradually decreased \cite{Hirsh}.
This change might have been caused by structural changes in US industries toward the digital economy.
\end{itemize}

The studies in the second stage can be summarized as follows.
\begin{itemize}
  \item In the US, the estimated long-run price-elasticity of the demand for commercial electricity varied from $-3.11$ to $-0.497$ depending on the states\cite{Bernstein2006}. The average elasticity was $-1.43$. Conversely, estimated long-run price-elasticity of the demand for residential electricity varied from $-0.618$ to $-0.058$. The average elasticity was $-0.282$.
  \item In East Asia, an analysis of the long-run price-elasticity of energy consumption, which is slightly different from electricity consumption, during 2000 and 2010 showed that the elasticity was $+0.212$, $-0.125$, $+0.212$, $-0.0837$, $-0.343$, $-0.997$, and $+ 4.17$ for Australia, Japan, China, India, Philippines, Thailand, and Singapore, respectively \cite{Phoumin}. The average was $-0.188$: here, Singapore was excluded from the average. 
\end{itemize}

The first stage studies show that correlations between the electricity demand and economic growth are very high. Therefore, we assume that the generic value of the correlation is approximately $1.0$.
The second stage studies show that the price-elasticity of the demand for electricity is a small negative value. Therefore, we assume that the generic value of the elasticity is approximately $-0.05$.
The susceptibility analyses of the electricity price on the economic growth indicate that a 1\% increase in electricity prices is associated with an approximately 0.05\% decrease in economic growth. 
This assumption was confirmed by a study on the impact of electricity price on economic growth in South Africa \cite{Khobai2017}. 
This study showed that a 1\% increase in electricity prices caused the economic growth to drop by 0.036\%.
We adopt the assumption that a 1\% increase in electricity prices is associated with an approximately 0.05\% decreases in economic growth for SIDS economies.
These analyses imply that the estimated LCOE increases of 223\% (=(397-123)/123) for the 100\% renewable power grid corresponds to an as high as 11\% decrease in economic growth. 
This decrease would result in significant negative impacts for a SIDS economy.

The estimated high LCOE is primarily due to the high initial investment cost of the electricity storage shown in Table \ref{tab:3}. 
Therefore, we need an innovation to permit the installation of electricity storage with a low initial investment cost.
For example, 
grid-to-vehicle (G2V) and vehicle-to-grid (V2G) technologies might provide an additional capacity of the electricity storage with a low cost \cite{Heinen2011}, or alternatively, 
excess solar PV supply could be used to produce hydrogen via the electrolysis of water \cite{Hosseini2016} \cite{IRENA2018}.
The produced hydrogen could then be stored in a hyperbaric chamber.
Hydrogen combustion in gas turbines would reduce fuel consumption, and as a result it would be possible to curtail $\rm{CO}_2$ emissions. 
This would make it possible to reduce the electricity storage capacity needed to balance the supply and demand. Consequently, a lower LCOE and therefore a higher economic growth would be attained with less $\rm{CO}_2$ emissions.

\section{Summary}
\label{sec:5}

We estimated the system-wise LCOE for a power grid with a high level of renewable energy to discuss how the estimated LCOE affects the economic growth in a SIDS economy.

Our grid model aims to describe the 1st phase of smart grid in order to estimate the electricity price for a power system with a high level of renewable energy integration in a SIDS economy.
We formulated our grid model as an optimization model.
The objective function was minimized under the global constraints of the demand-supply balance and some local constraints on the electricity storage. 

We quantitatively estimated a measure of the electricity price for power grids in a SIDS economy.
The estimation results of the system-wise LCOE using the grid model were discussed in terms of the nexus of energy, environment and economic growth on a SIDS economy.
While 100\% renewable energy is technologically possible with the usage of electricity storage, the estimated LCOE is as high as $397$\$/MWh.
This estimated LCOE is substantially higher than electricity prices for residential consumers in the US ($125$\$/MWh) and Japan ($253$\$/MWh).

Susceptibility analyses in South Africa, China, and the US show that a 1\% increase in electricity prices is associated with an approximately 0.05\% decrease in economic growth. 
These analyses imply that the estimated cost increase of 223\% with a 100\% renewable power grid corresponds to an as high as 11\% decrease in economic growth. 
This decrease in economic growth would result in significant negative impact on SIDS economies.

The obtained high LCOE is primarily due to the high initial investment cost of the electricity storage. 
Therefore we need an innovation to allow the use of electricity storage with a low initial investment cost.
For example, the excess solar PV supply could be used to produce hydrogen via the electrolysis of water.
The produced hydrogen could be stored in a hyperbaric chamber.
Hydrogen combustion in gas turbines would reduce the consumption of diesel fuel and as a result it would be possible to curtail $\rm{CO}_2$ emissions. 
This would make it possible to reduce the electricity storage capacity required to balance the supply and demand. Consequently, a lower LCOE and therefore higher economic growth would be attained with less $\rm{CO}_2$ emissions.

\begin{acknowledgements}
The author would like to thank Shigeru Kimura (Economic Research Institute for ASEAN and East Asia), Han Phoumin (Economic Research Institute for ASEAN and East Asia), Romeo Pacudan (Brunei National Energy Research Institute), Muhammad Nabih Fakhri Matussin (Brunei National Energy Research Institute), Majid Haji Sapar(Brunei National Energy Research Institute), Lam Kim Seong (Prudenergy Consulting Private Ltd.), Leong Siew Meng (Green Tech Solutions Inc.), and 
Takayuki Kusajima (Toyota Motor Corporation) for valuable comments.
\end{acknowledgements}

On behalf of all authors, the corresponding author states that there is no conflict of interest.


\begin{thebibliography}{}
%
\bibitem{Ohrlls2015}
UN-OHRLLS, ``SIDS in Numbers Climate Change Edition 2015'', (2015)
\bibitem{IPCC1990}
IPCC, ``IPCC 1990 First Assessment Report Overview Chapter'', (1990)
\bibitem{Sachs2015}
J.~D.~Sachs, ``The Age of Sustainable Development'', Columbia Univ Press, New York (2015)
\bibitem{Blanc2015}
D.~Le~Blanc, ``Towards integration at last? The sustainable development goals as a network of targets'', DESA Working Paper No. 141, March (2015)
\bibitem{Ikeda2014}
Y.~Ikeda and K.~Ogimoto, ``Cross-correlation of output fluctuation and system-balancing cost in photovoltaic integration'', Journal of Engineering 2014, pp.1-9, doi: 10.1049/joe.2014.0235 (2014)
\bibitem{Ikeda2015}
S.~Kimura, R.~Pacudan, H.~Phoumin, M.~N.~F.~Matussin, M.~H.~Sapar, L.~K.~Seong, L.~S.~Meng, T.~Kusajima, and Y.~Ikeda, ``Development of the Eco Town Model in the ASEAN Region through Adoption of Energy-Efficient Building Technologies, Sustainable Transport, and Smart Grids'', ERIA Research Project Report 2015, No. 20, chapter 4, March (2015)
\bibitem{Ikeda2018}
S.~Kimura, M.~N.~F.~Matussin, L.~S.~Meng, and Y.~Ikeda, ``Simulation Study on Energy Mix for Power Generation in Temburong Eco Town'', ERIA Research Project Report 2017, No.02, chapter 3, October (2018)
\bibitem{Chatzivasileiadis2013} 
S.~Chatzivasileiadis, D.~Ernst, G.~Andersson, ``The Global Grid'', Renewable Energy 57, pp. 372-383 (2013) 
\bibitem{IEA2015} 
International Energy Agency, ``Energy Prices and Taxes quarterly'', 4th Quarter (2015)
\bibitem{Lu}
W.~Lu, ``Electricity Consumption and Economic Growth: Evidence from 17 Taiwanese Industries'', Sustainability, 9, 50; doi:10.3390/su9010050 (2017)
\bibitem{Zhanga}
C.~Zhanga, K.~Zhoua, S.~Yanga, Z.~Shaoa, ``On electricity consumption and economic growth in China'', Renewable and Sustainable Energy Reviews 76, pp.353–368 (2017)
\bibitem{Hirsh}
R.~F.~Hirsh and J.~G.~Koomey, ``Electricity Consumption and Economic Growth: A New Relationship with Significant Consequences?'', The Electricity Journal, Vol. 28, Issue 9, pp.72-83, November (2015)
\bibitem{Bernstein2006}
M.~A.~Bernstein and J.~Griffin, ``Regional Differences in the Price-Elasticity of Demand for Energy'', NREL/SR-620-39512, February (2006)
\bibitem{Phoumin}
H.~Phoumin and S.~Kimura, ``Analysis on Price Elasticity of Energy Demand in East Asia: Empirical Evidence and Policy, Implications for ASEAN and East Asia'', ERIA-DP-2014-05, April (2014)
\bibitem{Khobai2017} 
H.~B.~Khobai, G.~Mugano, and P.~Le Roux, ``The Impact of Electricity Price on Economic Growth in South Africa'', International Journal of Energy Economics and Policy, 7(1), pp.108-116, (2017)
\bibitem{Heinen2011} 
S.~Heinen, D.~Elzinga, S-K.~Kim, Y.~Ikeda, ``Impact of smart grid technologies on peak load to 2050'', International Energy Agency, Working Paper, OECD Publishing, August (2011)
\bibitem{Hosseini2016} 
S.~E.~Hosseini, M.~A.~Wahid, ``Hydrogen production from renewable and sustainable energy resources: Promising green energy carrier for clean development'', Renewable and Sustainable Energy Reviews 57, pp.850-866 (2016)
\bibitem{IRENA2018}
IRENA, ``Hydrogen from renewable power: Technology outlook for the energy transition'', International Renewable Energy Agency, Abu Dhabi (2018)
%
\end{thebibliography}
\end{document}